\documentclass[conference]{IEEEtran}

\usepackage{cite}
\usepackage[cmex10]{amsmath}
\usepackage{graphicx}
\usepackage{picinpar}
\usepackage[cmex10]{amsmath}
\usepackage{amsmath,amsfonts,amssymb}
\usepackage{array}
\usepackage{mdwmath}
\usepackage{mdwtab}
\usepackage{eqparbox}
\usepackage{stfloats}
\usepackage{subfigure}
\usepackage{booktabs}
\usepackage{multirow}
\usepackage{etoolbox}
\usepackage{makecell}
\usepackage{color}
\usepackage{caption}

\usepackage[ruled]{algorithm2e}
\usepackage{fixltx2e}
\usepackage{bm}

\usepackage{fancyhdr} 

\begin{document}
\newtheorem{lemma}{Lemma}
\newtheorem{corol}{Corollary}
\newtheorem{theorem}{Theorem}
\newtheorem{proposition}{Proposition}
\newtheorem{definition}{Definition}
\newtheorem{remark}{Remark}
\newcommand{\e}{\begin{equation}}
\newcommand{\ee}{\end{equation}}
\newcommand{\eqn}{\begin{eqnarray}}
\newcommand{\eeqn}{\end{eqnarray}}


\title{Angle Estimation for Terahertz Ultra-Massive MIMO-Based Space-to-Air Communications} %

\vspace{-18mm}

\author{
\IEEEauthorblockN{Anwen Liao\IEEEauthorrefmark{1}, Zhen Gao\IEEEauthorrefmark{2}, Yang Yang\IEEEauthorrefmark{3}, Ha H. Nguyen\IEEEauthorrefmark{4}, Hua Wang\IEEEauthorrefmark{1}, and Hao Yin\IEEEauthorrefmark{5}}
\IEEEauthorblockA{\IEEEauthorrefmark{1}School of Information and Electronics, Beijing Institute of Technology, Beijing 100081, China}
\IEEEauthorblockA{\IEEEauthorrefmark{2}Advanced Research Institute of Multidisciplinary Science, Beijing Institute of Technology, Beijing 100081, China}
\IEEEauthorblockA{\IEEEauthorrefmark{3}School of Information and Communication Engineering, BUPT, Beijing 100876, China}
\IEEEauthorblockA{\IEEEauthorrefmark{4}Department of Electrical and Computer Engineering, University of Saskatchewan, Saskatoon, SK S7N 5A9, Canada}
\IEEEauthorblockA{\IEEEauthorrefmark{5}Institute of China Electronic System Engineering Corporation, Beijing 100141, China}
Email: gaozhen16@bit.edu.cn}
\vspace{-6mm}
\maketitle

\thispagestyle{fancy} 
\renewcommand{\headrulewidth}{0pt} 
\renewcommand{\footrulewidth}{0pt} 
\pagestyle{fancy}
\cfoot{\thepage} 

\begin{abstract}

 This paper investigates terahertz ultra-massive (UM)-MIMO-based angle estimation for space-to-air communications, which can solve the performance degradation problem caused by the dual delay-beam squint effects of terahertz UM-MIMO channels.
 Specifically, we first design a grouping true-time delay unit module that can significantly mitigate the impact of delay-beam squint effects to establish the space-to-air THz link.
 Based on the subarray selection scheme, the UM hybrid array can be equivalently considered as a low-dimensional fully-digital array, and then the fine estimates of azimuth/elevation angles at both UAVs and satellite can be separately acquired using the proposed prior-aided iterative angle estimation algorithm.
 The simulation results that close to Cram\'{e}r-Rao lower bounds verify the effectiveness of our solution.

\end{abstract}


\IEEEpeerreviewmaketitle

\vspace{-2mm}
\section{Introduction}\label{S1}
\vspace{-1.5mm}

 Terahertz (THz) communication is aimed at playing a vital role in the future sixth generation (6G) wireless communication networks \cite{Yang_Netw19}.
 To support up to terabit per second (Tbps) ultra-high peak data rate, THz-band, whose spectrum ranges from 0.1 to 10 THz, can offer tens of gigahertz (GHz) ultra-broadband bandwidths \cite{Akyildiz_CM18}.
 Meanwhile, ultra-massive multiple-input multiple-output (UM-MIMO)-based transceiver, which is equipped with tens of thousands of antennas \cite{Slim_JSAC19}, would be realized in THz communications.
 The UM-MIMO array of THz transceiver can utilize its large enough array aperture and beamforming techniques to effectively combat the severe path loss of THz signals and further extend the communication range \cite{Akyildiz_Access20}.
 Therefore, THz UM-MIMO technique is a favorable candidate for the unmanned aerial vehicles (UAVs) and satellite communications \cite{Akyildiz_Access20}.

 The reliable channel state information (CSI) acquisition is indispensable for the quality-of-service (QoS) of this space-to-air scenario \cite{KeML_TS20,KeML_JSAC21,MaXS_JSAC21}.
 However, the high-speed mobility of UAVs and satellites makes accurate CSI acquisition rather challenging.
 To acquire the dominant accurate channel parameters including the angles, Doppler shifts, and channel gains, some multi-stage channel estimation solutions were proposed in \cite{Qin_TVT18} for narrow-band millimeter-wave (mmWave) MIMO systems.
 In \cite{GaoXY_TVT17}, a priori-aided THz channel estimation scheme was proposed to predict and track the physical direction of line-of-sight (LoS) component of the time-varying massive MIMO channels in THz beamspace domain.

 However, due to the unprecedentedly ultra-broad band, ultra-large array aperture, and ultra-high velocity in the THz UM-MIMO based satellite communication systems, the aforementioned channel estimation schemes are difficult to be applied.
 Compared with the mmWave MIMO systems with limited bandwidth and aperture, the THz UM-MIMO based space-to-air channels present the unique \emph{triple delay-beam-Doppler squint effects}.
 Specifically, the THz UM-MIMO array equipped with tens of hundreds of antennas will suffer from different propagation delays at different antennas for the same impinging signal.
 These delay gaps may be as large as several symbol periods in ultra-broadband THz communications, which indicates the non-negligible inter-symbol-interference even for the LoS link.
 We can call this inevitable phenomenon in THz UM-MIMO systems as the {\emph{delay squint effect}}, which is also named as spatial-frequency wideband effects in \cite{GaoFF_TSP18,GaoFF_CM18}.
 Meanwhile, the {\emph{beam squint effect}}, in which the beam direction is a function of the operating frequency, can be further introduced by this delay squint effect.
 The large Doppler shift caused by the high-speed mobility in space-to-air THz communications is also frequency-dependent, i.e., {\emph{Doppler squint effect}}.
 The THz UM-MIMO based UAVs and satellite communication systems present {\emph{triple delay-beam-Doppler squint effects}}.
 Nevertheless, recent researches mainly focus on the impact of beam squint effect on mmWave systems \cite{GaoFF_TSP18,GaoFF_CM18,GaoFF_TSP20}.
 Consequently, an efficient channel parameter estimation solution is indispensable for THz UM-MIMO based space-to-air communications.

 In this paper, we mainly investigate the angle estimation for space-to-air LoS links connecting multiple UAVs and low earth orbit (LEO) satellite\footnote{Due to space limitation, we only estimate the azimuth and elevation angles at UAVs and satellite, while the remaining parts of channel estimation and tracking can be found in our work published in IEEE JSAC \cite{Liao_JSAC21}.}.
 To save space, we assume Doppler shifts can be well compensated using the positioning and flight posture information acquired by satellite communications, so that we just consider the dual delay-beam squint effects of space-to-air THz UM-MIMO channel (without Doppler squint effect) in this paper.
 Specifically, based on the rough angle estimates acquired from navigation information and positioning system, we first design a grouping true-time delay unit (GTTDU) module with low hardware cost to significantly mitigate the impact of delay-beam squint effects on both the transmitter and receiver, so as to establish the space-to-air THz link.
 After the link establishment, the UM hybrid array can be equivalently considered as a low-dimensional fully-digital array based on the subarray selection scheme.
 Then, the fine estimates of azimuth/elevation angles at both UAVs and satellite can be separately acquired using the proposed prior-aided iterative angle estimation algorithm.
 Simulations results have the good tightness with the Cram\'{e}r-Rao lower bounds (CRLBs) of azimuth/elevation angles, which testifies the good performance of the proposed solution.

\textit{Notations}:
 Italic boldface lower- and upper-case symbols denote column vectors and matrices, respectively.
 $(\cdot)^*$, $(\cdot)^{\rm T}$, and $(\cdot)^{\rm H}$ denote the conjugate, transpose, and Hermitian transpose operations, respectively.
 ${\| {\bm{a}} \|_2}$ is the ${\ell_2}$-norm of ${\bm{a}}$.
 $\otimes$ and $\circ$ denote the Kronecker and Hadamard product operations, respectively.
 $\bm{0}_n$ denotes the vector of size $n$ with all the elements being $0$.
 $|{\cal Q}|_c$ is the cardinality of the set ${\cal Q}$, and $\{{\cal Q}\}_n$ denotes the $n$th element of the ordered set ${\cal Q}$.
 $[\bm{a}]_{{\cal Q}}$ denotes the sub-vector containing the elements of $\bm{a}$ indexed in the ordered set ${\cal Q}$.
 $[\bm{a}]_m$ and $[\bm{A}]_{m,n}$ denotes the $m$th element of $\bm{a}$
 and the $m$th-row and the $n$th-column element of $\bm{A}$, respectively.
 Finally, $\mathbb{E}(\cdot)$ is the expectation of the argument.

\section{THz UM-MIMO Channel Model}\label{S2}

\begin{figure}[!tp]
\vspace{-2mm}
\begin{center}
 \includegraphics[width=0.85\columnwidth, keepaspectratio]{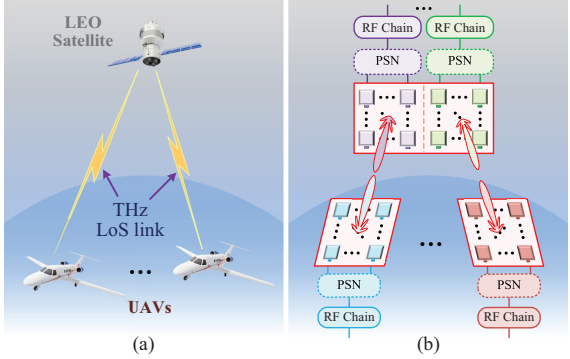}
\end{center}
\setlength{\abovecaptionskip}{-1mm}
\captionsetup{font = {footnotesize}, singlelinecheck = off, name = {Fig.}, labelsep = period}
\caption{(a) Typical space-to-air communication scenario among LEO satellite and multiple UAVs; and (b) the structure diagram of antenna arrays at transceiver, where UAVs that use analog beamforming communicate with satellite adopting subconnected PSN through respective LoS link.}
\label{FIG1}
\vspace{-4mm}
\end{figure}

 In this section, we will formulate the THz full-dimensional UM-MIMO channel model using uniform planar array (UPA), which involves azimuth and elevation angles \cite{Liao_Tcom19}.
 Fig.~\ref{FIG1}(a) depicts the specific scenario that $L$ UAVs communicate with a LEO satellite through respective THz LoS link.
 In Fig.~\ref{FIG1}(b), we can observe that the hybrid beamforming structure with a sub-connected phase shift network (PSN) can be  adopted in the transceiver at LEO satellite and UAVs \cite{Akyildiz_CM18}.
 The specific configurations of these antenna arrays are as follows.
 Defining $N_{\rm U}^{\rm h}$ and $N_{\rm U}^{\rm v}$ as the numbers of antennas in horizontal and vertical directions, respectively, and the total number of antennas at UAV arrays is $N_{\rm U}\! =\! N_{\rm U}^{\rm h}N_{\rm U}^{\rm v}$.
 Considering the sub-connected PSN at LEO satellite, ${\widetilde I}_{\rm S}^{\rm h}$
 ($M_{\rm S}^{\rm h}$) and ${\widetilde I}_{\rm S}^{\rm v}$ ($M_{\rm S}^{\rm v}$) can be defined as the numbers of subarrays (antennas within each subarray) in horizontal and vertical directions, respectively;
 while the numbers of antennas in horizontal and vertical directions of array are $N_{\rm S}^{\rm h}\! =\! {\widetilde I}_{\rm S}^{\rm h}M_{\rm S}^{\rm h}$ and
 $N_{\rm S}^{\rm v}\! =\! {\widetilde I}_{\rm S}^{\rm v}M_{\rm S}^{\rm v}$, respectively.
 Then, the total numbers of antennas in the whole antenna array (each subarray) is
 $N_{\rm S}\! =\! N_{\rm S}^{\rm h}N_{\rm S}^{\rm v}$ ($M_{\rm S}\! =\! M_{\rm S}^{\rm h}M_{\rm S}^{\rm v}$).
 Clearly, $L\! =\! {\widetilde I}_{\rm S}^{\rm h}{\widetilde I}_{\rm S}^{\rm v}$
 RF chains and only one RF chain are equipped at the LEO satellite and UAV, respectively,
 and each RF chain and the corresponding subarray mounted on LEO satellite are allocated to the specific UAV.

 To combat the multipath effect at the receiver of LEO satellite caused by multiple THz LoS links, the orthogonal frequency-division multiplexing (OFDM) technique with $K$ subcarriers will be applied to this space-to-air communication system.
 Specifically, the orthogonal frequency division multiple access (OFDMA) is utilized to split the pilot signals transmitted from different UAVs and improve the accuracy of angle estimation.
 Hence, considering the alternating subcarrier index allocation with equal intervals, $K$ subcarriers can be equally assigned to $L$ UAVs, and the ordered subcarrier index set corresponding to the $l$th UAV is ${\cal K}_l$ with $K_l = |{\cal K}_l|_c$.
 Due to the channel reciprocity of downlink (DL) and uplink (UL) in time division duplex (TDD) systems, we focus on the formulation of DL channel matrix next.
 The DL spatial-delay channel matrix at time $t$ corresponding to the $l$th UAV can be defined as $\bm{\bar H}_{{\rm DL},l}^{(t)}(\tau)\! \in\! \mathbb{C}^{N_{\rm U}\!\times\! N_{\rm S}}$, whose the ($n_{\rm U},n_{\rm S}$)th element, i.e., $[\bm{\bar H}_{{\rm DL},l}^{(t)}(\tau)]_{n_{\rm U},n_{\rm S}}$, can be expressed as
\begin{equation}\label{eq_H_tau_dl_mn} 
\vspace{-1mm}
 [\bm{\bar H}_{{\rm DL},l}^{(t)}(\tau)]_{n_{\rm U},n_{\rm S}} = \sqrt{G_l} \alpha_l e^{\textsf{j} {2\pi\psi_l t}}
 \delta\big( \tau - \tau_l - \underbrace{(\tau_l^{[n_{\rm U}]} + \tau_l^{[n_{\rm S}]})}_{\bm{Delay\,\,squint}} \big) , \nonumber
\vspace{-1mm}
\end{equation}
 where $1\! \le\! n_{\rm U}\! \le\! N_{\rm U}$, $1\! \le\! n_{\rm S}\! \le\! N_{\rm S}$,
 $G_l$ is the large-scale fading gain, $\alpha_l\! \sim\! {\cal CN}(0,\sigma_\alpha^2)$ denotes
 the channel gain\footnote{Due to the negligible attenuation of THz space-to-air communication links (e.g., atmospheric molecular absorption) in the stratosphere and above,
 the channel gain $\alpha_l$ can be modeled as a frequency flat coefficient.},
 $\psi_l\! =\! \underline{v}_l/\lambda_z$ is the Doppler shift with $\lambda_z$ and $\underline{v}_l$ being the carrier wavelength at central carrier frequency $f_z$ and the relative radial velocity, respectively,
 $\tau_l$ is the path delay,
 $\tau_l^{[n_{\rm U}]}$ and $\tau_l^{[n_{\rm S}]}$ denote the transmission delay for the $n_{\rm U}$th and $n_{\rm S}$th antennas at the UAV and satellite arrays, respectively,
 and $\delta(\cdot)$ is the Dirac impulse function.
 After some algebraic transformations, the DL channel matrix $\bm{H}_{{\rm DL},l}^{[m]}[k]$ in spatial-frequency domain at the $k$th subcarrier of the $n$th OFDM symbol for the $l$th UAV can be formulated as
\begin{align}\label{eq_H_k_ul_1} 
\vspace{-2mm}
 \bm{H}_{{\rm DL},l}^{[m]}[k] =& \sqrt{G_l} \alpha_l
 e^{\textsf{j} {2\pi \psi_l (m-1)T_{\rm sym}}} e^{-\textsf{j} {2\pi \left({\textstyle{k-1 \over K}}-{\textstyle{1 \over 2}}\right)f_s \tau_l}} \nonumber\\
 &\times \bm{A}_{{\rm DL},l}[k] ,
\end{align}
 where $f_s$ and $T_{\rm sym}$ are the system bandwidth and the duration time of an OFDM symbol, respectively, and $\bm{A}_{{\rm DL},l}[k]\! \in\! \mathbb{C}^{N_{\rm U}\!\times\! N_{\rm S}}$ denotes the DL array response matrix associated
 with the $l$th UAV, given by
\begin{align} 
\vspace{-1mm}
\bm{A}_{{\rm DL},l}[k] =& \underbrace{\left( \bm{a}_{\rm U}(\mu_l^{\rm U},\nu_l^{\rm U})
 \bm{a}^{\rm H}_{\rm S}(\mu_l^{\rm S},\nu_l^{\rm S}) \right)}_{\bm{A}_{{\rm DL},l}} \nonumber\\
 &\circ \underbrace{\left( \bm{\bar a}_{\rm U}(\mu_l^{\rm U},\nu_l^{\rm U},k) \bm{\bar a}^{\rm H}_{\rm S}(\mu_l^{\rm S},\nu_l^{\rm S},k)
 \right)}_{\bm{\bar{A}}_{{\rm DL},l}[k]\ {(\bm{Beam\,\,squint\,\,component})}} , \label{eq_A_ul_k1}
\end{align}
 where $\mu_l^{\rm U}\! =\! \pi\sin(\theta_l^{\rm U}) \cos(\varphi_l^{\rm U})$
 ($\mu_l^{\rm S}\! =\! \pi\sin(\theta_l^{\rm S})\cos(\varphi_l^{\rm S})$) and
 $\nu_l^{\rm U}\! =\! \pi\sin(\varphi_l^{\rm U})$ ($\nu_l^{\rm S} \! =\! \pi\sin(\varphi_l^{\rm S})$) that consider half-wavelength antenna spacing
 are the virtual angles at horizontal and vertical directions for the $l$th UAV (satellite), respectively,
 $\bm{A}_{{\rm DL},l}$ is the general DL array response matrix without beam squint effect,
 and $\bm{\bar{A}}_{{\rm DL},l}[k]$ is the array response squint matrix with beam squint effect.
 In (\ref{eq_A_ul_k1}), $\bm{a}_{\rm U}(\mu_l^{\rm U},\!\nu_l^{\rm U})\! =\!
 \bm{a}_{\rm v}(\nu_l^{\rm U},\!N_{\rm U}^{\rm v})\!\otimes\!
 \bm{a}_{\rm h}(\mu_l^{\rm U},\!N_{\rm U}^{\rm h})$
 and $\bm{a}_{\rm S}(\mu_l^{\rm S},\!\nu_l^{\rm S})\! =\!
 \bm{a}_{\rm v}(\nu_l^{\rm S},\!N_{\rm S}^{\rm v})\!\otimes\!
 \bm{a}_{\rm h}(\mu_l^{\rm S},\!N_{\rm S}^{\rm h})$
 are the regular array response vectors at the $l$th UAV and satellite \cite{Liao_JSAC21,Liao_Tcom19}, respectively,
 and $\bm{\bar a}_{\rm U}(\mu_l^{\rm U},\nu_l^{\rm U},k)\! =\!
 \bm{\bar a}_{\rm v}(\nu_l^{\rm U},N_{\rm U}^{\rm v},k)\!\otimes\!
 \bm{\bar a}_{\rm h}(\mu_l^{\rm U},N_{\rm U}^{\rm h},k)$
 and $\bm{\bar a}_{\rm S}(\mu_l^{\rm S},\nu_l^{\rm S},k)\! =\!
 \bm{\bar a}_{\rm v}(\nu_l^{\rm S},N_{\rm S}^{\rm v},k)\!\otimes\!
 \bm{\bar a}_{\rm h}(\mu_l^{\rm S},N_{\rm S}^{\rm h},k)$
 are the frequency-dependent array response squint vectors, respectively.
 Moreover, the expressions of 
 the horizontal/vertical steering vectors
 $\bm{a}_{\rm h}(\mu_l^{\rm U},N_{\rm U}^{\rm h})$ and
 $\bm{a}_{\rm v}(\nu_l^{\rm U},N_{\rm U}^{\rm v})$,
 and the horizontal/vertical steering squint vectors
 $\bm{\bar a}_{\rm h}(\mu_l^{\rm U},N_{\rm U}^{\rm h},k)$
 and $\bm{\bar a}_{\rm v}(\nu_l^{\rm U},N_{\rm U}^{\rm v},k)$
 can be found in \cite{Liao_JSAC21} for details.
 Note that the vectors at UAVs, i.e., $\bm{a}_{\rm h}(\mu_l^{\rm S},N_{\rm S}^{\rm h})$,
 $\bm{a}_{\rm v}(\nu_l^{\rm S},N_{\rm S}^{\rm v})$,
 $\bm{\bar a}_{\rm h}(\mu_l^{\rm S},N_{\rm S}^{\rm h},k)$,
 and $\bm{\bar a}_{\rm v}(\nu_l^{\rm S},N_{\rm S}^{\rm v},k)$,
 have the similar definitions and expressions.

 Similar to (\ref{eq_H_k_ul_1}), the UL spatial-frequency channel matrix at the $k$th subcarrier of the $m$th OFDM symbol for the satellite,
 denoted by $\bm{H}_{{\rm UL},l}^{[n]}[k]$, can be expressed as
\begin{equation}\label{eq_H_k_dl} 
 \bm{H}_{{\rm UL},l}^{[n]}[k] = \sqrt{G_l} \alpha_l e^{\textsf{j} {2\pi \psi_l (n-1)T_{\rm sym}}} \bm{A}_{{\rm UL},l}[k] .
\end{equation}

\vspace{-1mm}
\section{Proposed Angle Estimation Solution under Delay-Beam Squint Effects}\label{S3}
\vspace{-0.5mm}

 In this section, we propose a prior-aided iterative angle estimation solution for THz UM-MIMO based space-to-air communications, where some rough prior angles at UAVs and LEO satellite can be acquired from the positioning, flight speed and direction, and posture information for establishing the THz communication links.

\begin{figure}[!tp]
\vspace{-2mm}
\begin{center}
 \includegraphics[width=0.8\columnwidth, keepaspectratio]{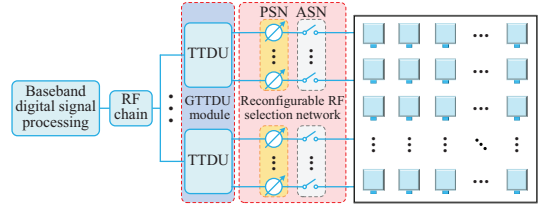}
\end{center}
\setlength{\abovecaptionskip}{-1mm}
\captionsetup{font = {footnotesize}, singlelinecheck = off, name = {Fig.}, labelsep = period}
\caption{The transceiver structure corresponding to one RF, where this RF chain connects with the antenna array via the GTTDU module and the reconfigurable RF selection network consisting of a sub-connected PSN and an ASN.}
\label{FIG2}
\vspace{-4mm}
\end{figure}

 To overcome the delay-beam squint effects of THz UM-MIMO array, we design a hardware-effective implementation of TTDU module, i.e., GTTDU module based transceiver structure as shown in Fig.~\ref{FIG2}, as the alternative of optimal TTDU module that is made up of numerous true-time delay units with each unit being assigned to its dedicated antenna \cite{LiYe_JSAC17}.
 From Fig.~\ref{FIG2}, we observe that except for the antenna array, this transceiver structure contains a GTTDU module and a reconfigurable RF selection network involving a subconnected
PSN and an antenna switching network (ASN), where this ASN can control the active or inactive state of the antenna elements to form different connection patterns of the RF selection network at the angle estimation stage.
 In this GTTDU module, a TTDU can be shared by a group of antennas and this imperfect hardware limitation can be handled by the subsequent signal processing algorithms well.

 Based on the prior information acquired from positioning systems, the rough estimates of
 azimuth and elevation angles at UAVs (satellite) can be defined as $\{ {\widetilde \theta}_l^{\rm U} \}_{l=1}^{L}$
 ($\{ {\widetilde \theta}_l^{\rm S} \}_{l=1}^{L}$) and $\{ {\widetilde \varphi}_l^{\rm U} \}_{l=1}^{L}$
 ($\{ {\widetilde \varphi}_l^{\rm S} \}_{l=1}^{L}$), respectively, and the corresponding
 horizontally and vertically virtual angles are
 $\{ {\widetilde \mu}_l^{\rm U} \}_{l=1}^{L}$ ($\{ {\widetilde \mu}_l^{\rm S} \}_{l=1}^{L}$)
 and $\{ {\widetilde \nu}_l^{\rm U} \}_{l=1}^{L}$ ($\{ {\widetilde \nu}_l^{\rm S} \}_{l=1}^{L}$), respectively.
 According to $\bm{H}_{{\rm DL},l}^{[m]}[k]$ in (\ref{eq_H_k_ul_1}), we present the DL spatial-frequency channel matrix $\bm{\widetilde H}_{{\rm DL},l}^{[m]}[k]$ compensated by ideal TTDU module as
\begin{align}\label{eq_H_ttdu_k_dl} 
\vspace{-2mm}
 \bm{\widetilde H}_{{\rm DL},l}^{[m]}[k] =& \sqrt{G_l} \alpha_l
 e^{\textsf{j} {2\pi \psi_l (m-1)T_{\rm sym}}} e^{-\textsf{j} {2\pi \left({\textstyle{k-1 \over K}}-{\textstyle{1 \over 2}}\right)f_s \tau_l}} \nonumber\\
 &\times \bm{\widetilde A}_{{\rm DL},l}[k] ,
\end{align}
 in which
\begin{equation}\label{eq_A_ttdu_dl_k} 
\vspace{-1mm}
 \bm{\widetilde A}_{{\rm DL},l}[k]\! =\! \bm{A}_{{\rm DL},l}[k] \circ
 {\underbrace{\left( \bm{\bar a}_{\rm U}({\widetilde \mu}_l^{\rm U},{\widetilde \nu}_l^{\rm U},k)
 \bm{\bar a}^{\rm H}_{\rm S}({\widetilde \mu}_l^{\rm S},{\widetilde \nu}_l^{\rm S},k) \right)}_{\widetilde{\bm{\bar{A}}}_{{\rm DL},l}[k]}}^* .
 \vspace{-1mm}
\end{equation}
 By comparing $\widetilde{\bm{\bar{A}}}_{{\rm DL},l}[k]$ in (\ref{eq_A_ttdu_dl_k}) and $\bm{\bar{A}}_{{\rm DL},l}[k]$ in (\ref{eq_A_ul_k1}), we can find that the beam squint effect part can be perfectly eliminated if we can acquire the perfect angle information, i.e., $\widetilde{\bm{\bar{A}}}_{{\rm DL},l}[k]\! =\! \bm{\bar{A}}_{{\rm DL},l}[k]$ and then $\bm{\widetilde A}_{{\rm DL},l}[k]\! =\! \bm{A}_{{\rm DL},l}$ when ${\widetilde \mu}_l^{\rm U}\! =\! \mu_l^{\rm U}$, ${\widetilde \nu}_l^{\rm U}\! =\! \nu_l^{\rm U}$, ${\widetilde \mu}_l^{\rm S}\! =\! \mu_l^{\rm S}$, and ${\widetilde \nu}_l^{\rm S}\! =\! \nu_l^{\rm S}$.
 Moreover, the compensated UL spatial-frequency channel matrix $\bm{\widetilde H}_{{\rm UL},l}^{[n]}[k]$ has the similar expressions, which are omitted for simplicity.
 The practical DL/UL channel matrices compensated by the GTTDU module can be derived from (\ref{eq_H_ttdu_k_dl}) and (\ref{eq_A_ttdu_dl_k}).

 At the angle estimation stage, the azimuth/elevation angles at UAVs can be estimated in DL, while those angles at satellite are acquired in UL.

\vspace{-1mm}
\subsection{Fine Angle Estimation at UAVs}\label{S3.1}
\vspace{-0.5mm}

\begin{figure*}[!tp]
\vspace{-2mm}
\begin{center}
 \includegraphics[width=1.3\columnwidth, keepaspectratio]{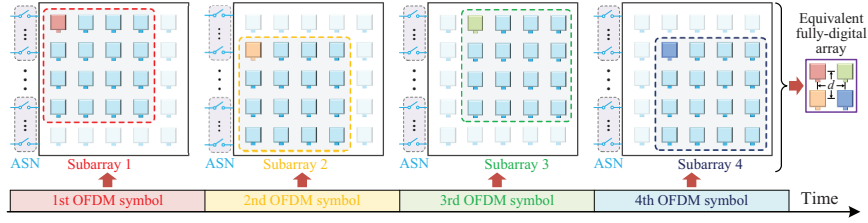}
\end{center}
\setlength{\abovecaptionskip}{-0.2mm}
\captionsetup{font = {footnotesize}, singlelinecheck = off, name = {Fig.}, labelsep = period}
\caption{Schematic diagram of subarray selection scheme at the angle estimation stage, where the different antenna connection patterns can be formed by controlling the ASN of the reconfigurable RF selection network.}
\label{FIG3}
\vspace{-4mm}
\end{figure*}

 Due to the limited valid observation at the UAVs, it is necessary to accumulate multiple OFDM symbols for estimating the azimuth and elevation angles.
 Meanwhile, the transmitted signals can be compensated the priori Doppler shifts well, that is, the compensated channels will be slow time-varying.
 By altering the RF connection pattern at the antenna array of UAV, the received signals that adopt different selected subarrays will differ by one envisaged phase value if the transceiver has the same configuration, and these organized phase differences can construct the array response vector of low-dimensional fully-digital array.
 In Fig.~\ref{FIG3}, we take the UPA of size $5\!\times\! 5$ as an example.
 By controlling the reconfigurable RF selection network $4$ subarrays of size $4\! \times\! 4$ in $4$ successive OFDM symbols can be selected to form the array response vector of equivalent fully-digital array with size of $2\! \times\! 2$ with the critical antenna spacing $d$.

 Specifically, $I_{\rm U}$ OFDM symbols are used to estimate the angles at UAVs, where each OFDM symbol adopts a selected subarray (corresponding to the dedicated RF connection pattern).
 By employing the rough angle estimates at satellite and UAVs, we can first design the analog precoding and combining vectors, i.e., $\bm{p}_{{\rm RF},l}$ and $\bm{q}_{{\rm RF},l}^{[m]}$ for $1\! \le\! l\! \le\! L$, $1\! \le\! m\! \le\! I_{\rm U}$.
 For $\bm{p}_{{\rm RF},l}$, initialize $\bm{p}_{{\rm RF},l}$ as $\bm{p}_{{\rm RF},l}\! =\! \bm{0}_{N_{\rm S}}$, and then let $[\bm{p}_{{\rm RF},l}]_{{\cal I}_{{\rm S},l}}\! =\! \textstyle{1 \over \sqrt{M_{\rm S}}} [\bm{a}_{\rm S}({\widetilde \mu}_l^{\rm S},{\widetilde \nu}_l^{\rm S})]_{{\cal I}_{{\rm S},l}}$,
 where ${\cal I}_{{\rm S},l}$ with $M_{\rm S}\! =\! |{\cal I}_{{\rm S},l}|_c$ denotes the antenna index of subarray assigned to the $l$th UAV due to each subarray at satellite only communicating with its corresponding UAV as shown in Fig.~\ref{FIG1}(b).
 To design $\{\bm{q}_{{\rm RF},l}^{[m]}\}_{m=1}^{I_{\rm U}}$, we can partition the UM-MIMO array at UAV into $I_{\rm U}\! =\! I_{\rm U}^{\rm h}I_{\rm U}^{\rm v}$ smaller subarrays for yielding the array response vector of equivalent low-dimensional fully-digital array with size of $I_{\rm U}^{\rm h}\! \times\! I_{\rm U}^{\rm v}$.
 The sizes of these smaller subarrays are ${\bar M}_{\rm U}^{\rm h}\! \times\! {\bar M}_{\rm U}^{\rm v}$ (${\bar M}_{\rm U}^{\rm h}\! =\! N_{\rm U}^{\rm h}\! -\! I_{\rm U}^{\rm h}\! +\! 1$ and ${\bar M}_{\rm U}^{\rm v}\! =\! N_{\rm U}^{\rm v}\! -\! I_{\rm U}^{\rm v}\! +\! 1$) and their number of antennas is ${\bar M}_{\rm U}\! =\! {\bar M}_{\rm U}^{\rm h}{\bar M}_{\rm U}^{\rm v}$.
 For $1\! \le\! i_{\rm U}^{\rm h}\! \le\! I_{\rm U}^{\rm h}$ and $1\! \le\! i_{\rm U}^{\rm v}\! \le\! I_{\rm U}^{\rm v}$, by defining $m\! =\! (i_{\rm U}^{\rm v}-1)I_{\rm U}^{\rm h} \! +\! i_{\rm U}^{\rm h}$ with $i_{\rm U}^{\rm h}$ and $i_{\rm U}^{\rm v}$ being the ($i_{\rm U}^{\rm h},i_{\rm U}^{\rm v}$)th subarray, respectively,
 we denote the antenna index of the selected $m$th subarray corresponding to the $m$th OFDM symbol as ${\cal I}_{\rm U}^{[m]}$, where ${\bar M}_{\rm U}\! =\! |{\cal I}_{\rm U}^{[m]}|_c$
 So $\bm{q}_{{\rm RF},l}^{[m]}$ can be also initialized as $\bm{q}_{{\rm RF},l}^{[m]}\! =\! \bm{0}_{N_{\rm U}}$,
 and then let $[\bm{q}_{{\rm RF},l}^{[m]}]_{{\cal I}_{\rm U}^{[m]}}\! =\! \textstyle{1 \over \sqrt{{\bar M}_{\rm U}}} [\bm{a}_{\rm U}({\widetilde \mu}_l^{\rm U},{\widetilde \nu}_l^{\rm U})]_{{\cal I}_{\rm U}^{[1]}}$ for $1\! \le\! m\! \le\! I_{\rm U}$.

 The received signal $y_{{\rm DL},l}^{[m]}[k_l]$ at the $k_l$th subcarrier of the $m$th OFDM symbol transmitted by the $l$th UAV is given by
\begin{equation}\label{eq_y_ukl} 
\vspace{-2mm}
 y_{{\rm DL},l}^{[m]}[k_l] = \sqrt{P_l} (\bm{q}_{{\rm RF},l}^{[m]})^{\rm H}
 \bm{\widetilde H}_{{\rm DL},l}^{'[m]}[k_l] \bm{p}_{{\rm RF},l} s_{{\rm DL},l}^{[m]}[k_l] + n_{{\rm DL},l}^{[m]}[k_l] ,
\end{equation}
 where $k_l\! \in\! {\cal K}_l$, $1\! \le\! m\! \le\! I_{\rm U}$,
 $\bm{\widetilde H}_{{\rm DL},l}^{'[m]}[k_l]$ is the channel matrix compensated by the Doppler shifts and the GTTDU module, and $s_{{\rm DL},l}^{[m]}[k_l]$ and $n_{{\rm DL},l}^{[m]}[k_l]$ are the transmitted pilot signal and noise, respectively.
 By collecting the received signals at $K_l$ subcarriers as $\bm{y}_{{\rm DL},l}^{[m]}\! \in\! \mathbb{C}^{K_l}$, we have
\begin{equation}\label{eq_y_ul_vec1} 
 \bm{y}_{{\rm DL},l}^{[m]}\! =\! \sqrt{P_lG_l} \alpha_l (\bm{q}_{{\rm RF},l}^{[m]})^{\rm H} \bm{A}_{{\rm DL},l} \bm{p}_{{\rm RF},l} \bm{s}_{{\rm DL},l}^{[m]}
 \circ \bm{\widetilde y}_{{\rm DL},l}^{[m]}\! +\! \bm{n}_{{\rm DL},l}^{[m]} ,
\end{equation}
 where $\bm{s}_{{\rm DL},l}^{[m]}\! =\! \left[ s_{{\rm DL},l}^{[m]}[\{{\cal K}_l\}_1]\! \cdots\! s_{{\rm DL},l}^{[m]}[\{{\cal K}_l\}_{K_l}]\right]^{\rm T}\! \in\! \mathbb{C}^{K_l}$,
 $\bm{\widetilde y}_{{\rm DL},l}^{[m]}$ is the error vector including the residual beam squint caused by inaccurate prior information.
 Moreover, the same transmitted pilot signals are adopted for $I_{\rm U}$ OFDM symbol, i.e., $\bm{s}_{{\rm DL},l}\! =\! \bm{s}_{{\rm DL},l}^{[m]}$ for $1\! \le\! m\! \le\! I_{\rm U}$.
 By stacking $\{\bm{y}_{{\rm DL},l}^{[m]}\}_{m=1}^{I_{\rm U}}$ received from $I_{\rm U}$ OFDM symbols, we can obtain $\bm{Y}_{{\rm DL},l}\! =\! \left[ \bm{y}_{{\rm DL},l}^{[1]}\cdots \bm{y}_{{\rm DL},l}^{[I_{\rm U}]} \right]^{\rm T}\! \in\! \mathbb{C}^{I_{\rm U}\! \times\! K_l}$, i.e.,
\begin{align}\label{eq_Y_ul_angle1} 
\vspace{-1mm}
 \bm{Y}_{{\rm DL},l} =& \sqrt{P_lG_l} \alpha_l \left( \bm{Q}_{{\rm RF},l}^{\rm H} \bm{A}_{{\rm DL},l} \bm{p}_{{\rm RF},l} \bm{s}_{{\rm DL},l}^{\rm T} \right)
 \circ \bm{\widetilde Y}_{{\rm DL},l} \nonumber\\
 &+ \bm{N}_{{\rm DL},l} ,
\end{align}
 where $\bm{\widetilde Y}_{{\rm DL},l}\! =\! \left[ \bm{\widetilde y}_{{\rm DL},l}^{[1]}\! \cdots\! \bm{\widetilde y}_{{\rm DL},l}^{[I_{\rm U}]} \right]\! \in\! \mathbb{C}^{I_{\rm U}\! \times\! K_l}$
 and $\bm{Q}_{{\rm RF},l}\! =\! \left[ \bm{q}_{{\rm RF},l}^{[1]}\! \cdots\!
 \bm{q}_{{\rm RF},l}^{[I_{\rm U}]} \right]\! \in\! \mathbb{C}^{N_{\rm U}\!\times\! I_{\rm U}}$
 are the residual beam squint and the analog combining matrices, respectively.
 Based on this analog combining matrix $\bm{Q}_{{\rm RF},l}$, we can extract the array response
 vector of equivalent low-dimensional fully-digital  to estimate the angles at UAVs using the robust array signal processing techniques.
 Specifically, compared with
 $(\bm{q}_{{\rm RF},l}^{[1]})^{\rm H}\bm{a}_{\rm U}(\mu_l^{\rm U},\nu_l^{\rm U})$ for
 $m\! =\! 1$ in (\ref{eq_y_ul_vec1}), $(\bm{q}_{{\rm RF},l}^{[m]})^{\rm H}\bm{a}_{\rm U}(\mu_l^{\rm U},\nu_l^{\rm U})$
 is multiplied by an extra phase shift $e^{\textsf{j} {\left( {(i_{\rm U}^{\rm h}-1)\mu_l^{\rm U}\! +\! (i_{\rm U}^{\rm v}-1)\nu_l^{\rm U}} \right)}}$ for $m\! =\! (i_{\rm U}^{\rm v}-1)I_{\rm U}^{\rm h} \! +\! i_{\rm U}^{\rm h}$ and $2\! \le\! m\! \le\! I_{\rm U}$.
 Obviously, the effective array response vector of equivalent fully-digital array with size of $I_{\rm U}^{\rm h}\! \times\! I_{\rm U}^{\rm v}$ can be constituted by these regular phase shifts, that is, ${\bm{\bar{\bar a}}}_{\rm U}
 (\mu_l^{\rm U},\nu_l^{\rm U})\! =\! \bm{a}_{\rm v}(\nu_l^{\rm U},I_{\rm U}^{\rm v})\!\otimes\!
 \bm{a}_{\rm h}(\mu_l^{\rm U},I_{\rm U}^{\rm h})\! \in\! \mathbb{C}^{I_{\rm U}}$.
 Thus, $\bm{Y}_{{\rm DL},l}$ in (\ref{eq_Y_ul_angle1}) can be then rewritten as
\begin{equation}\label{eq_Y_ul_angle2} 
\vspace{-1mm}
 \bm{Y}_{{\rm DL},l} = \gamma_{{\rm DL},l} \left( {\bm{\bar{\bar a}}}_{\rm U}(\mu_l^{\rm U},\nu_l^{\rm U})
 \bm{s}_{{\rm DL},l}^{\rm T} \right) \circ \bm{\widetilde Y}_{{\rm DL},l} + \bm{N}_{{\rm DL},l} ,
\end{equation}
 where $\gamma_{{\rm DL},l}\! =\! \sqrt{P_lG_l} \alpha_l (\bm{q}_{{\rm RF},l}^{[1]})^{\rm H}
 \bm{A}_{{\rm DL},l} \bm{p}_{{\rm RF},l}$ denotes the beam-aligned effective channel gain.

\SetAlCapFnt{\normalsize}
\SetAlCapNameFnt{\normalsize}
\SetAlFnt{\small}
\begin{algorithm}[tp!]
\caption{Prior-Aided Iterative Angle Estimation}\label{ALG1}
\LinesNumbered
\KwIn{Rough virtual angles $\{ {\widetilde \mu}_l^{\rm U},{\widetilde \nu}_l^{\rm U},{\widetilde \mu}_l^{\rm S},{\widetilde \nu}_l^{\rm S} \}_{l=1}^L$, received signal $\bm{Y}_{{\rm DL},l}$, and maximum iterations $i_{\rm U}^{\rm max}$}
\KwOut{Estimated azimuth/elevation angles $\{ {\widehat \theta}_l^{\rm U},{\widehat \varphi}_l^{\rm U} \}$ and virtual angles $\{ {\widehat \mu}_l^{\rm U},{\widehat \nu}_l^{\rm U} \}$ for $1\! \le\! l\! \le\! L$}
\For{$i_{\rm U}\! =\! 1,\! \cdots\!, i_{\rm U}^{\rm max}$}
{ \eIf{$i_{\rm U}\! =\! 1$}
{ {Apply TDU-ESPRIT algorithm to $\bm{Y}_{{\rm DL},l}$}\;
 {Obtain angle estimates of first iteration as $\{ {\widehat \theta}_l^{(i_{\rm U})},{\widehat \varphi}_l^{(i_{\rm U})} \}$ and $\{ {\widehat \mu}_l^{(i_{\rm U})},{\widehat \nu}_l^{(i_{\rm U})} \}$}\;
}
{ {Design compensation matrix $\bm{\widetilde Y}_{{\rm DL},l}^{(i_{\rm U}-1)}$, whose $k_l$th column $\bm{\widetilde y}_{{\rm DL},l}^{(i_{\rm U}-1)}[k_l]$ is shown in (\ref{eq_y_tilde_ul_pie})}\;
 {Obtain compensated matrix $\bm{Y}_{{\rm DL},l}^{(i_{\rm U})}\! =\! \left(\bm{\widetilde Y}_{{\rm DL},l}^{(i_{\rm U}-1)}\right)^{*} \circ \bm{Y}_{{\rm DL},l}$ in (\ref{eq_Y_ul_pie})}\;
 {Apply TDU-ESPRIT algorithm to $\bm{Y}_{{\rm DL},l}^{(i_{\rm U})}$}\;
 {Obtain angle estimates of $i_{\rm U}$th iteration as $\{ {\widehat \theta}_l^{(i_{\rm U})},{\widehat \varphi}_l^{(i_{\rm U})} \}$ and $\{ {\widehat \mu}_l^{(i_{\rm U})},{\widehat \nu}_l^{(i_{\rm U})} \}$}\;
}
}
{\bf Return}: ${\widehat \theta}_l^{\rm U}\! =\! {\widehat \theta}_l^{(i_{\rm U}^{\rm max})}$, ${\widehat \varphi}_l^{\rm U}\! =\!  {\widehat \varphi}_l^{(i_{\rm U}^{\rm max})}$, ${\widehat \mu}_l^{\rm U}\! =\! {\widehat \mu}_l^{(i_{\rm U}^{\rm max})}$, and ${\widehat \nu}_l^{\rm U}\! =\! {\widehat \nu}_l^{(i_{\rm U}^{\rm max})}$ for $1\! \le\! l\! \le\! L$
\end{algorithm}

 Next, a prior-aided iterative angle estimation  is proposed as follows.
 By applying the two-dimensional unitary ESPRIT (TDU-ESPRIT) algorithm \cite{Liao_Tcom19} to
 the received signal matrix $\bm{Y}_{{\rm DL},l}$ in (\ref{eq_Y_ul_angle2}), the proposed algorithm at the first iteration $i_{\rm U}\! =\! 1$ estimates the azimuth and elevation angles (the corresponding horizontally and vertically virtual
 angles) at the $l$th UAV, i.e., ${\widehat \theta}_l^{(i_{\rm U})}$
 and ${\widehat \varphi}_l^{(i_{\rm U})}$ (${\widehat \mu}_l^{(i_{\rm U})}$ and ${\widehat \nu}_l^{(i_{\rm U})}$) for $1\! \le\! l\! \le\! L$.
 Furthermore, to minimize the impact of $\bm{\widetilde Y}_{{\rm DL},l}$ on (\ref{eq_Y_ul_angle2}), the estimated angles above can be utilized to iteratively compensate $\bm{Y}_{{\rm DL},l}$ at the subsequent iterations (i.e., $i_{\rm U}\! \ge\! 2$), so as to acquire more accurate angle estimates.
 To be specific, according to the roughly priori virtual angle estimates ${\widetilde \mu}_l^{\rm U}$ and ${\widetilde \nu}_l^{\rm U}$,
 and ${\widehat \mu}_l^{(i_{\rm U}-1)}$ and ${\widehat \nu}_l^{(i_{\rm U}-1)}$ estimated
 at the $(i_{\rm U}\! -\! 1)$th iteration,
 the compensation matrix at the $i_{\rm U}$th iteration can be defined as
 $\bm{\widetilde Y}_{{\rm DL},l}^{(i_{\rm U}-1)}\! =\! \left[ \bm{\widetilde y}_{{\rm DL},l}^{(i_{\rm U}-1)}[\{{\cal K}_l\}_1]\! \cdots\! \bm{\widetilde y}_{{\rm DL},l}^{(i_{\rm U}-1)}[\{{\cal K}_l\}_{K_l}] \right]$,
 whose the $k_l$th column $\bm{\widetilde y}_{{\rm DL},l}^{(i_{\rm U}-1)}[k_l]\! \in\! \mathbb{C}^{I_{\rm U}}$ can be expressed as
\begin{align}\label{eq_y_tilde_ul_pie} 
\vspace{-1mm}
 \bm{\widetilde y}_{{\rm DL},l}^{(i_{\rm U}-1)}[k_l] =& \left( \bm{\bar a}_{\rm v}({\widetilde \nu}_l^{\rm U},I_{\rm U}^{\rm v},k_l)
 \otimes \bm{\bar a}_{\rm h}({\widetilde \mu}_l^{\rm U},I_{\rm U}^{\rm h},k_l) \right)^* \nonumber\\
 &\circ\! \left( \bm{\bar a}_{\rm v}({\widehat \nu}_l^{(i_{\rm U}-1)},I_{\rm U}^{\rm v},k_l)
 \otimes \bm{\bar a}_{\rm h}({\widehat \mu}_l^{(i_{\rm U}-1)},I_{\rm U}^{\rm h},k_l) \right) .
\end{align}
 The processed matrix $\bm{Y}_{{\rm DL},l}^{(i_{\rm U})}\! =\! \left(\bm{\widetilde Y}_{{\rm DL},l}^{(i_{\rm U}-1)}\right)^{*} \circ \bm{Y}_{{\rm DL},l}$ compensated by $\bm{\widetilde Y}_{{\rm DL},l}^{(i_{\rm U}-1)}$ is
\begin{align}\label{eq_Y_ul_pie} 
\vspace{-1mm}
 \bm{Y}_{{\rm DL},l}^{(i_{\rm U})} =\ &\gamma_{{\rm DL},l} \left( {\bm{\bar{\bar a}}}_{\rm U}(\mu_l^{\rm U},\nu_l^{\rm U})
 \bm{s}_{{\rm DL},l}^{\rm T} \right) \nonumber\\
 &\circ \left( \bm{\widetilde Y}_{{\rm DL},l} \circ
 \left(\bm{\widetilde Y}_{{\rm DL},l}^{(i_{\rm U}-1)}\right)^{*} \right) + \bm{N}_{{\rm DL},l}^{(i_{\rm U})} .
\end{align}
 To obtain the more accurate angle estimates, we can apply the TDU-ESPRIT algorithm to those matrices $\{ \bm{Y}_{{\rm DL},l}^{(i_{\rm U})} \}_{l=1}^L$ until the maximum iterations $i_{\rm U}^{\rm max}$ is reached, i.e., $i_{\rm U}\! =\! i_{\rm U}^{\rm max}$
 and the estimated azimuth/elevation angles at UAVs and the corresponding virtual angles are ${\widehat \theta}_l^{\rm U}\! =\! {\widehat \theta}_l^{(i_{\rm U}^{\rm max})}$,
 ${\widehat \varphi}_l^{\rm U}\! =\!  {\widehat \varphi}_l^{(i_{\rm U}^{\rm max})}$,
 ${\widehat \mu}_l^{\rm U}\! =\! {\widehat \mu}_l^{(i_{\rm U}^{\rm max})}$, and
 ${\widehat \nu}_l^{\rm U}\! =\! {\widehat \nu}_l^{(i_{\rm U}^{\rm max})}$ for $1\! \le\! l\! \le\! L$.
 
 \textbf{Algorithm~\ref{ALG1}} summarizes the proposed prior-aided iterative angle estimation procedure above, which can address the beam squint effect well.

\subsection{Fine Angle Estimation at Satellite}\label{S3.2}

 According to the TDD channel reciprocity, estimating the fine angle estimates at satellite in UL is similar to the DL angle estimation of UAVs in Section~\ref{S3.1}.
 At this stage, the fine angles estimated at UAVs above can replace the originally rough angle estimates to refine the GTTDU modules at UAVs and design the analog precoding vectors at UAVs for beam alignment with improved receive signal-to-noise ratio (SNR).
 
 To be specific, considering $I_{\rm S}\! =\! I_{\rm S}^{\rm h}I_{\rm S}^{\rm v}$
 OFDM symbols to estimate the fine azimuth/elevation angles at satellite,
 we can obtain the low-dimensional fully-digital array with size of $I_{\rm S}^{\rm h}\! \times\! I_{\rm S}^{\rm v}$.
 By employing the estimated $\{ {\widehat \mu}_l^{\rm U},{\widehat \nu}_l^{\rm U} \}_{l=1}^L$,
 we design the analog precoding vector as $\bm{f}_{{\rm RF},l}\! =\!
 \bm{a}_{\rm U}({\widehat \mu}_l^{\rm U},{\widehat \nu}_l^{\rm U})$ for $1\! \le\! l\! \le\! L$.
 According to the reconfigurable RF selection network, we denote the selected antenna index in the $n$th OFDM sysmbol at the $l$th aircraft subarray as ${\cal I}_{{\rm S},l}^{[n]}$, where ${\bar M}_{\rm S}\! =\! |{\cal I}_{{\rm S},l}^{[n]}|_c$.
 The analog combining vector can be then initialized as $\bm{w}_{{\rm RF},l}^{[n]}\! =\! \bm{0}_{N_{\rm S}}$, and let $[\bm{w}_{{\rm RF},l}^{[n]}]_{{\cal I}_{{\rm S},l}^{[n]}}\! =\! \textstyle{1 \over \sqrt{{\bar M}_{\rm S}}}
 [\bm{a}_{\rm S}({\widetilde \mu}_l^{\rm S},{\widetilde \nu}_l^{\rm S})]_{{\cal I}_{{\rm S},l}^{[1]}}$,
 for $1\! \le\! n\! \le\! I_{\rm S}$, $1\! \le\! l\! \le\! L$.

 Similar to (\ref{eq_y_ukl})-(\ref{eq_Y_ul_angle2}), the received signal $\bm{Y}_{{\rm UL},l}$ can be also obtained, where the details can be found in \cite{Liao_JSAC21} due to the limited space.
 By replacing the input parameters
 $\{ {\widetilde \mu}_l^{\rm U},{\widetilde \nu}_l^{\rm U},\bm{Y}_{{\rm DL},l},i_{\rm U},i_{\rm U}^{\rm max} \}$ for UAVs with the corresponding parameters
 $\{ {\widehat \mu}_l^{\rm U},{\widehat \nu}_l^{\rm U},\bm{Y}_{{\rm UL},l},i_{\rm S},i_{\rm S}^{\rm max} \}$ for satellite, we can utilize the proposed prior-aided iterative angle estimation algorithm in \textbf{Algorithm~\ref{ALG1}} to acquire more accurate estimates of azimuth and elevation angles and the corresponding virtual angles at satellite, i.e., ${\widehat \theta}_l^{\rm S}\! =\! {\widehat \theta}_l^{(i_{\rm S}^{\rm max})}$,
 ${\widehat \varphi}_l^{\rm S}\! =\!  {\widehat \varphi}_l^{(i_{\rm S}^{\rm max})}$,
 ${\widehat \mu}_l^{\rm S}\! =\! {\widehat \mu}_l^{(i_{\rm S}^{\rm max})}$, and
 ${\widehat \nu}_l^{\rm S}\! =\! {\widehat \nu}_l^{(i_{\rm S}^{\rm max})}$ for $1\! \le\! l\! \le\! L$.

\subsection{Computational Complexity Analysis}\label{S3.3}

 The computational complexity of the proposed angle estimation solution consists of the acquisition of azimuth/elevation angles at UAVs and satellite using TDU-ESPRIT algorithm.
 Their total computational complexity is $\textsf{O}\left(i_{\rm U}^{\rm max}LI_{\rm U}K_l\! +\! i_{\rm S}^{\rm max}LI_{\rm S}K_l\right)$.
 Since the effective low-dimensional signals at the UAVs and satellite are utilized to estimate the angles, the computational complexity of the proposed angle estimation solution is in polynomial time even although the THz UM-MIMO arrays are equipped with tens of thousands of antennas at UAVs and satellite.

\section{Numerical Evaluation}\label{S4}

 In this section, we evaluate the performance of the proposed angle estimation solution for THz UM-MIMO-based space-to-air communications.
 Without loss of generality, the LEO satellite serves $L\! =\! 2$ UAVs, where the vertical distance between satellite and UAVs is $200$ kilometer and these two UAVs are randomly appeared in a horizontal circular plane with radius $R_{\rm a}\! =\! 50\, {\rm km}$.
 The relative radial velocity among UAVs and satellite is $200$ meter per second.
 Other simulation parameter settings are shown in Table~\ref{TAB1}.
 Moreover, we define $\sigma_\alpha^2/\sigma_n^2$ with $\sigma_n^2$ being the noise variance as the transmitted SNR in DL and UL.
 The performance of angle estimation is evaluated using the root mean square error (RMSE) metric given by ${\mathrm{RMSE}}_{\bm{x}}\! =\! \sqrt{ \mathbb{E}\left({\textstyle{1 \over L}}
 \| \bm{x}\! -\! \bm{\widehat x} \|_2^2\right) }$, where $\bm{x}\! \in\! \mathbb{R}^{L}$ and
 $\bm{\widehat x}$ represent the true and the estimated angle vectors with $[\bm{x}]_l$ being $\theta_l^{\rm U}$, $\varphi_l^{\rm U}$, $\theta_l^{\rm S}$, or $\varphi_l^{\rm S}$.
 To evaluate the estimation performance, the CRLBs serve as the lower bounds of angle estimation \cite{Paulraj_TSP98}.

\begin{table}[!tp]
\renewcommand\arraystretch{1.4}
\captionsetup{labelsep = period}
\centering
\caption{Simulation Parameter Settings}
\label{TAB1}
\vspace{-4mm}
\begin{center}
\begin{tabular}{l!{\vrule width0.8pt}l}
\Xhline{0.8pt}
 {\textbf{Parameter}} & {\textbf{Value}} \\
\Xhline{0.8pt}
 $f_z$ ($f_s$) & $0.1$\,THz ($1$\,GHz) \\
\hline
 $N_{\rm U}^{\rm h}$, $N_{\rm U}^{\rm v}$, $M_{\rm S}^{\rm h}$, $M_{\rm S}^{\rm v}$ & $200$ \\
\hline
 ${\widetilde I}_{\rm S}^{\rm h}$ (${\widetilde I}_{\rm S}^{\rm v}$) & $1$ ($2$) \\
\hline
 $I_{\rm U}^{\rm h}$, $I_{\rm U}^{\rm v}$, $I_{\rm S}^{\rm h}$, $I_{\rm S}^{\rm v}$ & $5$ \\
\hline
 ${\widetilde M}_{\rm U}^{\rm h}$, ${\widetilde M}_{\rm U}^{\rm v}$, ${\widetilde M}_{\rm S}^{\rm h}$, ${\widetilde M}_{\rm S}^{\rm v}$ & $5$ \\
\hline
 $K$ (cyclic prefix $N_{\rm cp}$) & $2048$ ($128$) \\
\hline
 $\{ \theta_l^{\rm U},\! \varphi_l^{\rm U},\! \theta_l^{\rm S},\!
 \varphi_l^{\rm S} \}_{l=1}^L$ & $-60^\circ - 60^\circ$ \\
\hline
 $\{ \tau_l \}_{l=1}^L$ ($\{ \alpha_l \}_{l=1}^L$) & ${\cal U}[0, N_{\rm cp}T_s]$ (${\cal CN}(0,1)$) \\
\hline
 Maximum offset of rough angle estimates & $\pm\, 5^\circ$ \\
\Xhline{0.8pt}
\end{tabular}
\end{center}
\vspace{-6mm}
\end{table}

\begin{figure*}[!tp]
\vspace{-2mm}
\captionsetup{singlelinecheck = off, justification = raggedright, font={footnotesize}, name = {Fig.}, labelsep = period}
\captionsetup[subfigure]{singlelinecheck = on, justification = raggedright, font={footnotesize}}
\centering
\subfigure{
\label{FIG4(a)}
\includegraphics[width=3.25in]{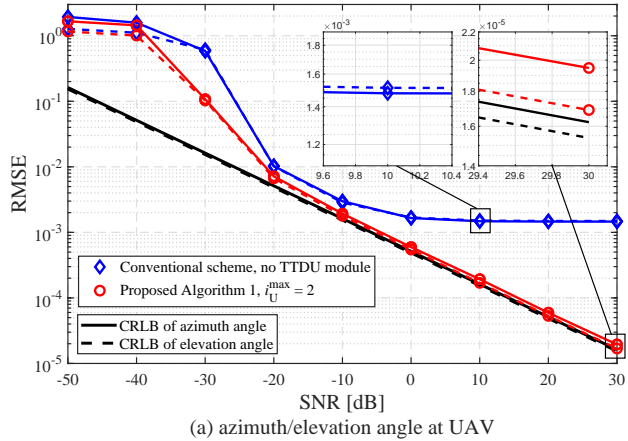}
}
\subfigure{
\label{FIG4(b)}
\includegraphics[width=3.25in]{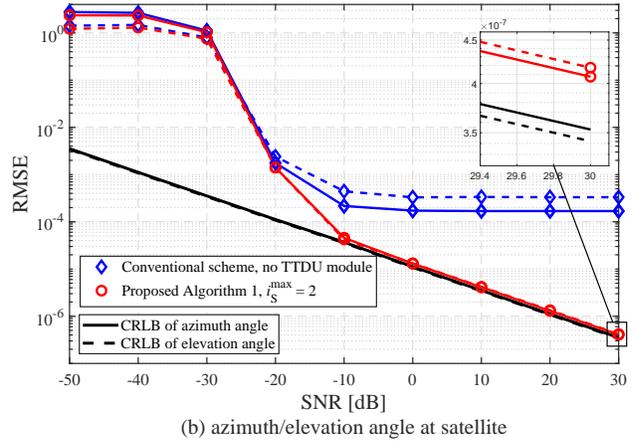}
}
\setlength{\abovecaptionskip}{-0.05mm}
\caption{RMSE comparison of azimuth/elevation angles $\{ \theta_l^{\rm U},\varphi_l^{\rm U},\theta_l^{\rm S},\varphi_l^{\rm S} \}_{l=1}^L$ at the UAVs and satellite.}
\label{FIG4}
\vspace{-2.5mm}
\end{figure*}

 Fig.~\ref{FIG4} compares the RMSE performance of the proposed fine angle estimation solution for the azimuth/elevation angles at the UAVs and satellite.
 Here, the label ``conventional scheme, no TTDU module'' indicates the transceiver without considering TTDU module directly applies the conventional TDU-ESPRIT algorithm to estimate angles.
 From Figs.~\ref{FIG4(a)} and \ref{FIG4(b)}, it can be seen that the RMSE curves of ``proposed algorithm 1'' for the estimated azimuth/elevation angles at the UAVs and satellite are very close to the CRLBs at medium-to-high SNRs, where the proposed \textbf{Algorithm~\ref{ALG1}} just needs $i_{\rm U}^{\rm max}\! =\! 2$ and $i_{\rm S}^{\rm max}\! =\! 2$  iterations to achieve the upper-bound performance.
 If the beam squint effect is not well handled as ``conventional scheme'' with ``no TTDU module'', its performance of angle estimation will suffer from the obvious RMSE floor at medium-to-high SNRs.
 Furthermore, comparing Figs.~\ref{FIG4(a)} and \ref{FIG4(b)}, it can be observed that more accurate angle estimates can be obtained at satellite.
 This is because the angle estimation at satellite employs more accurate angles estimated at UAVs to obtain the larger beam alignment gain than the angle estimation at UAVs.

\vspace{-1.0mm}
\section{Conclusions}
\vspace{-0.5mm}

 In this paper, we have proposed a prior-aided iterative angle estimation solution, which can effectively solve the dual delay-beam squint effects of THz UM-MIMO-based space-to-air channel.
 By designing the GTTDU module, we first significantly mitigate the impact of delay-beam squint effects to establish the space-to-air THz link.
 On this basis, the UM hybrid array can be equivalently considered as a low-dimensional fully-digital array using the subarray selection scheme.
 Then, we can separately acquire the fine estimates of azimuth/elevation angles at UAVs and satellite by utilizing the proposed prior-aided iterative angle estimation algorithm.
 Simulation results have confirmed that the proposed solution can obtain the good angle estimation performance.

\vspace{-1.0mm}
\section*{Acknowledgement}
\vspace{-1.0mm}
This work was supported by the International Flagship Partnership Research Grant (IFPRG), the National Natural Science Foundation of China (NSFC) under Grant Nos. 62071044 and 61801035, and the Beijing Natural Science Foundation (BJNSF) under Grant No. L182024.

\end{document}